# The Clausius inequality: implications for nonequilibrium thermodynamic steady states with NEMD corroboration

Christopher Gunaseelan Jesudason
Chemical Physics Division, Chemistry Department, University of Malaya
50603 Kuala Lumpur, Malaysia   email:jesu@um.edu.my

**Abstract:** The Clausius inequality for closed systems was deduced from the Riemann integration of closed Carnot cycle loops for irreversible transitions. The corresponding inequalities for the recently developed open system Carnot cycles are derived here and their properties indicate that no new nonequilibrium entropy results from them as has been proposed over the decades. It is proven that a sequence of points along a non-equilibrium state space must have excess variables augmenting those for the equilibrium situation, thereby proving that the Principle of Local Equilibrium (PLE) used extensively to describe nonequilibrium systems is only an approximation. To demonstrate the breakdown of the Principle, verification of the theorem, and the presence of new phenomena, an ab initio steady state simulation of a simple dimer reaction $2A \leftrightarrow A_2$ under nonequilibrium conditions was performed and the results compared to the equilibrium conditions to show the regime of breakdown of the Principle. The novel dimer reaction presented has a cyclical pathway found in many natural processes, such as laser photochemistry. It is suggested that such reaction pathways can exist and awaits experimental verification. A new algorithm to conserve momentum and energy at the potential switches of the dimer is applied effectively. New atomic and molecular flux flows not present under equilibrium conditions are shown to exist, where the net rate of reaction along the cell is not zero but small, unlike the zero rate equilibrium requirement, leading to the flux presence. The equilibrium constant is also shifted from the value at thermodynamical equilibrium. Unless reinterpreted differently, it is shown that the so-called Curie symmetry principle, fundamental in deciding on the feasibility of flux-force couplings in both theory and experiment in all physical theories, cannot apply to such experimental results. Hence, a far greater freedom in selecting appropriate force-flux couplings is indicated than would be the case if the Curie principle, as commonly interpreted is adopted.

## 1 Introduction to open Carnot cycle analysis

To consider multi-component thermodynamical systems, an open Carnot sytem operating between two reservoirs at temperatures $T_1$ and $T_2$ are given and its properties during summation limits are used [1]. A typical cycle is given in Fig. 1, where the state space $\Sigma = \{\mathbf{P}, \mathbf{V}, \mathbf{m}\}$ denote the system intensive and extensive variables and the mass amounts present respectively; $\mathbf{m} = \{m_1, m_2 ... m_i ... m_n\}$ is the set of 'pure' substances injected into the open system; $\mathbf{m} \cup \mathbf{V}$ are the extensive quantities such that $-\mathbf{P} \cdot (\delta \mathbf{V} + \delta \mathbf{m})$ represents the work gained by the system for arbitrary displacements $(\delta \mathbf{V} + \delta \mathbf{m})$.

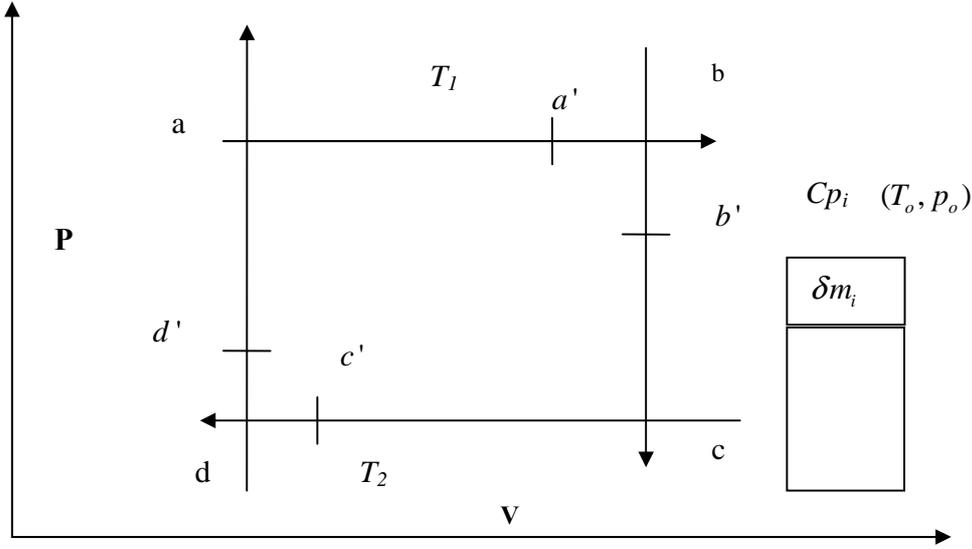

Figure 1: A system taken through a reversible cycle along abcda in multidimensional space $(\mathbf{P},\mathbf{V},\mathbf{m})$ with mass exchange $\delta m_i$.

Thermodynamical points $a,b,c,d$ also represents the state variables $\Sigma$ for those labeled states. Two basic cycles are depicted, $C_{iso}$ where mass in inserted at the isothermal segments of the Carnot cycle and $C_{adia}$, where they are inserted without heat exchange at the adiabatic portions. For cycles $C_{iso}$, $\delta Q^{T_1}_{a-a',sys}$ is the heat absorbed without mass change along a-a', masses $\delta m_i$ are injected reversibly via semi-permeable membranes at constant temperature along a'-b accompanied by the simultaneous exchange of heat $-\delta Q^{T_2}_{c-c',sys}$, and the heat absorbed along $cc'$ is $-\delta Q^{T_2}_{cc'}$ and when mass is extracted at c'd the heat gained by diathermal heat transfer through the boundary is $-\delta Q^{T_1}_{inj,i}$ via the system isothermal boundaries, whereas for $C_{adia}$ cycles there is no temperature control for the system during reversible exchange of masses. The temperatures along isotherms $ab$ and $cd$ is $T_1$ and $T_2$ respectively; and any of these cycles, heat is exchanged with the thermal reservoirs labeled $T_1$ and $T_2$ held respectively at the temperatures $T_1$ and $T_2$ there is also the work done to transport material $i$ from the supply cell $Cp_i$ at standard state (ss) $(T_o,p_o)$ to the surface of the reactor cell $\Delta W_i^{form}$, where the element $\delta m_i$ is in equilibrium with the reactor cell through the semi-permeable membrane with pressure-temperature variables $(p_i,T)$. The normal work terms in a system transition are denoted $\delta W$ for transitions along state $\Sigma$ when there is no mass exchange (as in a normal Carnot engine), with superscripts and subscripts indicating the transition coordinates, $\delta W_{inj}$ denotes the work of injecting the specified material into the reactor cell. Then it has been shown [1] that the total work done on the environment by the system $\Delta W_{tot,iso}$ and by the heat pumps, all of which work cyclically along $a \to b \to c \to d.$ are as follows

$$\Delta W_{tot,iso} = -\delta W^{T_1}_{aa'} - \delta W^{T_2}_{cc'} - \delta W_{bc} - \delta W_{da} + \delta V^{(T_1,p_{i,\mathbf{b}})}_{ss}$$
$$-\delta V^{(T_2,p_{i,\mathbf{d}})}_{ss} + \Delta W_i^{form}(p_{i,\mathbf{b}},T_1) - \Delta W_i^{form}(p_{i,\mathbf{d}},T_2)$$
$$-\delta W_{inj}(a'b) + \delta W_{inj}(c'd).$$

(1)

Total heat *lost* at $T_1$ reservoir $\Delta Q_{1,tot,iso}$ is

$$\Delta Q_{1,tot,iso} = \delta Q_{aa',sys}^{T_1} + \delta Q_{inj,i}^{T_1} + T_1 \delta m_i \Delta \mathscr{S}_{i,ss}^{(T_1,p_i)}. \quad (2)$$

Total heat *gained* at $T_2$, $\Delta Q_{2,tot,iso}$ is

$$\Delta Q_{2,tot,iso} = \delta Q_{c'c,sys}^{T_2} - \delta Q_{inj,i}^{T_2} + T_2 \delta m_i \Delta \mathscr{S}_{i,ss}^{(T_2,p_{i,c'd})}$$
$$= \delta Q_{cc',sys}^{T_2} - \delta Q_{inj,i}^{T_2} + T_2 \delta m_i \Delta \mathscr{S}_{i,ss}^{(T_2,p_{i,c'd})} \quad (3)$$

where $\Delta \mathscr{S}_{i,ss}^{(T,p_i)} = \left( \int_{ss}^{(p_i,T)} \frac{dQ_i}{T} \right)_i$ is the convected entropy per unit mass for substance $i$ with heat contribution deriving from the indicated reservoir. More explicitly, $\delta \mathscr{W}_{ss}^{(T_1,p_i)} = \Delta \mathscr{W}_{i,ss}^{(T,p_i)} \delta m_i = T \delta m_i \Delta \mathscr{S}_{i,ss}^{(T,p_i)} - \Delta Q_{i,ss}^{(T,p_i)} \delta m_i$ is the work done on the environment for the pumping mechanism of species $i$ with mass $\delta m_i$ where $\Delta Q_{i,ss} = \int_{ss}^{(p_i,T)} dQ$ and $\Delta Q_i^{(p_i,T)} = T \left( \int_{ss}^{(p_i,T)} \frac{dQ_i}{T} \right) \delta m_i = T \delta m_i \Delta \mathscr{S}_{i,ss}^{(T,p_i)}$.

The corresponding results for an elementary cycle for the injection and extraction of mass with the above notation for an adiabatic $C_{adia}$ cycle where mass is extracted or injected is (subscripted adia. refers to the $C_{adia}$ cycle and other subscripts refer to the path or state)

$$\Delta W_{tot,adia} = -\delta W_{ab}^T - \delta W_{cd}^{T_2} - \delta W_{b'c} - \delta W_{d'a} + \delta \mathscr{V}_{ss}^{(T_1,p_{i,\mathbf{bb'}})}$$
$$-\delta \mathscr{V}_{ss}^{(T_2,p_{i,\mathbf{d'}})} + \Delta W_i^{form}(p_{i,\mathbf{b}},T_1) - \Delta W_i^{form}(p_{i,\mathbf{d}},T_2).$$

(4)

$$\Delta Q_{1,tot,adia} = \delta Q_{ab,sys}^{T_1} + \delta Q_{inj,i}^{T_1} + T_1 \delta m_i \Delta \mathscr{S}_{i,ss}^{(T_1,p_i)}. \quad (5)$$

$$\Delta Q_{2,tot,adia} = \delta Q_{dc,sys}^{T_2} + \delta Q_{inj,i}^{T_2} + T_2 \delta m_i \mathscr{S}_{i,ss}^{(T_2,p_i(c'd))}. \quad (6)$$

We note that there can be no heat absorption about an adiabatic segment. For a mixed elementary loop cycle $C_{comb}$ where mass is injected(extracted) at an isothermal pathway e.g. ab and extracted(injected) at the adiabatic pathway e.g. da, we still derive $\Delta W_{tot,comb}, \Delta Q_{1,tot,comb}$ and $\Delta Q_{2,tot,comb}$, as above. For the above cycles (denoted $C$) operating between temperature points 1 and 2, and for general transitions the following (Theorems 1-3 below) has been proven [1].

**Theorem 1.** Each elementary cycle $C$ fulfills $\dfrac{\Delta Q_{1,tot,C}}{T_1} - \dfrac{\Delta Q_{2,tot,C}}{T_2} = 0$ for optimized Carnot trajectories.

**Theorem 2.** A perfect differential $d\mathscr{S} = \dfrac{dQ_{tot}}{T}$ for the state function $\mathscr{S}$ exists, given by

$$d\mathscr{S} = \frac{dQ_{dia}}{T} + \sum_{i=1}^n \Delta \mathscr{S}_{i,ss}^{(T_1,p_i)} dm_i \text{ where } dQ_{tot} = dQ_{dia} + T \sum_{i=1}^n \Delta \mathscr{S}_{i,ss}^{(T_1,p_i)} dm_i \text{ and }$$

$$dQ_{dia} = dQ_{sys} + \sum_{i=1}^n dQ_{inj,i}.$$

The total reversible diathermal heat transfer increment $dQ_{dia}$ consists of $dQ_{sys}$ which is the heat absorption by the diathermal boundary of the system when there is no mass transfer taking place and to $dQ_{inj,i}$ when there is a transfer of substance $i$ of amount $dm_i$.

**Theorem 3.** There exists an entropic function of state $\mathscr{S}_{dia}$ with differential given by $d\mathscr{S}_{dia} = \dfrac{dQ_{dia}}{T}$ where $Q_{dia}$ is a function representing the total heat absorption of the system through a diathermal boundary as above. This state function is local in the sense that it refers to quantities measured at the system during system transitions.

The traditional Clausius Inequality was derived using some of the following well founded Axioms, which will be adhered to here.

**Axiom 1.** (Kelvin-Clausius) It is impossible to construct an engine working in a cycle, which will produce no other effect than the transfer of heat from a cooler body to a hotter one.

**Axiom 2.** (Riemann summation) The $\Sigma$ variables vary continuously in the system transitions and are differentiable and in particular the **m** variables vary continuously even within a single open Carnot loop $j$ which is an element in the Riemann sum, so that no finite instances of measured mass injections/extractions can occur for any $j$ loop.

**Axiom 3.** (First Law) The energy functions have perfect total derivatives i.e. they are state functions.

**Axiom 4.** (Clausius inequality assumption) Irreversible pathways traverse a pathway along a sequence of points exactly that of a reversible pathway where the $\Sigma$ variables are concerned in an arbitrary circular, but yield different diathermal heat absorption increments for each element j of the Riemann summation.

**Remark 1.** If the above did not obtain, it would be impossible to form Riemann sums correlating a reversible and irreversible Carnot cycle about an infinitesimally small loop to deduce the Clausius inequality.

**Theorem 4.** 1. The Clausius inequality corresponding to Theorem 2 for the nonlocal heat increment is $\oint \dfrac{dQ_{tot}}{T} \leq 0$.

**Proof.** The optimized elementary reversible Carnot cycle yields from Theorem 1 zero for fixed temperature of reservoirs, so that the non-optimized cycle can only be $\dfrac{\Delta Q_{1,tot,C}}{T_1} - \dfrac{\Delta Q_{2,tot,C}}{T_2} \leq 0$ for any elementary loop (Axiom 4.). Taking N cycles ($N \to \infty$) to complete the Riemann sum for any loop yields $\underset{N \to \infty}{Lim} \sum_{j=1}^{N} \left( \dfrac{\Delta Q_{1,tot,C,j}}{T_j} - \dfrac{\Delta Q_{2,tot,C,j}}{T_j} \right) \leq 0 \Rightarrow \oint \dfrac{dQ_{tot}}{T} \leq 0$ □

The next proof requires Axioms 1 and 2. The $dQ_{tot}$ heat increment is a combination of local and non-local heat terms.

**Theorem 5.** The Clausius Inequality corresponding to Theorem 3 is $\oint \dfrac{dQ_{dia}}{T} \leq 0$.

**Proof.** One must consider whether optimized processes are involved or not in the partitioned local heat increment $dQ_{dia}$ which is not similar to the non-local total heat increment $dQ_{tot}$, where an optimized process is involved to derive an inequality by examining the energy transfer terms with the supply cells $Cp,i$. Let $U_{i,um}$ be the intensive energy variable relative to the supply cell to extract unit mass of substance $i$ to state $\Sigma$ of the primary cell (system) at equilibrium with it through a semi-permeable membrane, so that the actual energy increment transferred to the surface prior to any work being done on it is $d\mathscr{U}_i = U_{i,um}(\Sigma)dm_i$ for substance $i$. Gibbs' integration leads to the total energy (which has a perfect differential for it is a state function) of superficial transfer $\mathscr{U}$ being $\mathscr{U} = \sum_{i=1}^{m} m_i U_{i,um}(\Sigma) = \sum_{i=1}^{m} m_i \mathscr{U}_i$ where $\mathscr{U}_i = m_i U_{i,um}(\Sigma)$. Since $d\mathscr{U} = \sum_{i=1}^{m} dm_i U_{i,um}(\Sigma)$ for an $m$-substance system, another Gibbs-Duhem type equation $\sum_{i=1}^{m} m_i d(U_{i,um}(\Sigma)) = 0$ is proven to exist here. For any $j^{th}$ open Carnot engine, the energy associated with the convected mass prior to injection is written $\sum_{i=1}^{m} dm_{i,j} U_{i,um,j}(\Sigma)$ where the sign of $dm_{i,j}$ determines injection or extraction. For this cycle, the work energy to inject (extract) $dm_k$ by the environment is $-W_{inj,k,j}(\Sigma)dm_k$ and the external work done by the machine is $W_{ext}^j = \oint \mathbf{P.dV} + \sum_{k=1}^{m} \oint_{\partial m_k} W_{inj,k,j}(\Sigma)dm_k$. In one cycle, the whole system is returned to the original state. From Axiom 2, for the $j$ th cycle define $\Delta\mathscr{U}_j = \oint_{\mathbf{m}} \sum_{i=1}^{m} dm_i U_{i,um}(\Sigma)$ and $Q_1^j, Q_2^j$ are the total diathermal heat absorptions in this cycle at temperatures $T_1, T_2$ respectively. Clearly, $\Delta\mathscr{U}_j = 0$ (Axiom 3). The connection of the $Q$'s to the diathermal entropy is $\oint d\mathscr{S} = \oint \frac{dQ_{dia}}{T} = \lim_{N \to \infty} \sum_{j=1}^{N} \left( \frac{Q_1^j}{T_1} - \frac{Q_2^j}{T_2} \right) = 0$. The conservation of energy for each cycle loop j (Axiom 3) is

$$\Delta\mathscr{U}_j + Q_1^j - Q_2^j - W_{ext}^j = 0. \qquad (7)$$

As $j \to \infty$ to complete a loop in $\Sigma$ space, $\oint d\mathscr{U} = \sum_{j=1}^{N} \Delta\mathscr{U}_j = 0$, and so from (7)

$$\lim_{N \to \infty} \sum_{j=1}^{N} \Delta\mathscr{U}_j + Q_1^j - Q_2^j - W_{ext}^j = 0 \Rightarrow \oint dQ_{dia} = \oint dW_{ext} = \Delta W_{ext} \qquad (8)$$

where $\Delta W_{ext}$, the total work done by the system in any arbitrary loop in $\Sigma$ space, equals the total reversible diathermal heat absorbed. The above was derived implicitly in [1] when deriving a Principle of Correlation. For each $j$ cycle, the virtual closed Carnot engine has work done $dW_{vir}^j$, the maximum given by $dW_{vir}^{j,\max} = Q_1^j \left( \frac{T_1^j - T_2^j}{T_1^j} \right)$ from which energy conservation gives $\frac{Q_1^j}{Q_2^j} = \frac{T_1}{T_2}$. Now, from Axiom 4, the temperatures for the virtual and real system are the same at the diathermal ends, and if $Q_1^j$ is fixed, then so is $Q_2^j$. However if this $Q_2^j$ denoted $Q_2^j{'}$ is different from the open system $Q_2^j$, then for this cycle running one engine against another would violate Axiom 1 for the reversible situation, hence they must be the same. Comparing with (7) yields $Q_1^j - Q_2^j = dW_{vir}^{j,\max} = dW_{ext}^j - \Delta \mathscr{U}_j$, so that summing over all $j$ loops yield for the reversible case

$$\oint dW_{vir} = \Delta W_{vir} = \oint dW_{ext} = \Delta W_{ext} \qquad (9)$$

where the work done by a closed Carnot engine is precisely that of the open system; this is the Principle of Correlation developed previously [1]. If the efficiency of the irreversible engine exceeded that of the closed Carnot cycle at a particular $j$ cycle, then as before running this system coupled to a reversible system would lead to a violation of Axiom 1. Hence if the actual work $W_{ext}^j$ is less than the maximum reversible work for a particular $j$ cycle $W_{ext}^j$, then $W_{ext}^j < W_{ext}^{j,\max} \Rightarrow \left( \dfrac{Q_1^j}{T_1} - \dfrac{Q_2^j}{T_2} \right) < 0$ and summing this as $j \to \infty$ (as in Theorem 4) over a closed loop leads to $\oint \dfrac{dQ_{dia}}{T} \leq 0$ □

It is noted that two types of Clausius inequalities are derived, where the heat term in Theorem 5 is local. The above results contradicts Bhalekar's proof [2 a] that the inequality does not exist for open systems and also to an extent the rebuttal [2 b] since no heat terms are discriminated here and no proven state functions were apparently derived.

## 2 Applications of theorems

2.1 Non-existence of excess entropy function of state

Two types of heat increment terms are used here, both satisfying Clausius-type inequalities denoted $dQ_{[q]}: q = \{adia, tot\}$ where the adia subscript refers to the adiabatic heat increment, and the other to the non-local total heat increment as discussed in Theorem 4 and 5, with the associated entropy forms $d\mathscr{S}_{[q]}$ forms respectively, where $d\mathscr{S}_{[tot]} \equiv d\mathscr{S}$ of Theorem 2. There have been attempts by a tradition set by Eu [3 a-j] and others [4 a-c] over the last two decades to derive a new entropy form based on heat compensation from the Clausius inequality with supposed applications despite not having a proper characterization of the heat terms used for the Clausius inequalities, and of the entropy forms. Most irreversible systems are open in nature, but the theorem was derived initially using the traditional closed Clausius loop, and later with Gibbs' thermodynamical assumptions. A proof for closed systems that these new entropy forms cannot obtain was provided recently [5], and since the mathematical structure of both open and closed systems are the same as shown above, the results given in [5] may be generalized for the 2 $q$ forms $q = \{adia, tot\}$ given here (above). For any two points $A, B \subset \Sigma$ define two pathways connecting these points in a continuous curve, $P_{AB}$ and $P'_{AB}$ which are along a reversible and irreversible pathway respectively. Writing the Clausius integral as $-N_{[q]}$ and integrating between A and B between two paths $P_{AB}$ and $P'_{AB}$ about a closed loop yields $-N_{[q]} = \oint_{irrev} \dfrac{dQ_{[q]}}{T} \leq 0$ or

$N_{[q]} = \Delta \mathscr{S}_{[q]} - \int_{A,irr}^{B} \dfrac{dQ_{[q]}[P'_{AB}]}{T} \geq 0$ where $\Delta \mathscr{S}_{[q]} = \int_{A,rev}^{B} \dfrac{dQ_{[q]}[P_{AB}]}{T}$ is the reversible entropy change between A and B.

**Lemma 1.** The variable $N_{[q]}$ must be a functional of the variable A,B and path $P'_{AB}$, i.e. $N_{[q]} = N_{[q]}(A, B, P'_{AB})$.

**Proof.** Since $\Delta \mathscr{S}_{[q]}$ is the integral of a perfect differential, it is a function of the endpoints of the integral, and the irreversible integration along $P'_{AB}$ is path dependent, hence the result □

Defining $N_{[q]} = \oint dN_{[q]}$, the above notation leads to $\oint \left( \frac{dQ_{[q]}}{T} + dN_{[q]} \right) = 0$, suggestive of a perfect differential $d\Sigma'_{[q]} = \frac{dQ_{[q]}}{T} + dN_{[q]}$. The following determines the issue.

**Theorem 6.** The differential $d\Sigma'_{[q]} = \frac{dQ_{[q]}}{T} + dN_{[q]}$ is not exact and is nonlocal.

**Proof.** Replacing the $N, Q, S, \Sigma$ variables in reference [5] (Secs. 3 (a) and 3(b)) by $N_{[q]}, Q_{[q]}, \mathscr{S}_{[q]}, \Sigma'_{[q]}$ and repeating the argument for our general system here, where the algebra is isomorphous because of the proof of the Clausius inequality here for general systems, the above theorem follows since it is proven there [eq. 13(b), 5] that $d\Sigma = \frac{dQ}{T} + dN$ in the notation of reference [5] *is not an exact differential and furthermore is nonlocal* □

Eu has offered a reconciliation in that if his entropy is viewed as coincident with the nonequilibrium entropy, then there is convergence between E.I.T [4 c,d] and his approach [p.771;3 c]

2.2 Considerations in principle of local equilibrium (PLE)

This principle [p.23; 6 a: p.7; 6 b: p.7, 6 c] states that a nonequilibirum steady state system may be locally described by variables that describe an equilibrium system, and that the heat transfer and other flux terms arise from the gradients in the equilibrium variables across the entire system. Some have suggested this principle to be very strong on the basis of a restricted simulation and other studies of particles with an interparticle potential [7 a-c] . Theorem 7 below shows that there are augmenting variables to irreversible systems without contradicting Axiom 4, which is used to derive the theorem. Implications are then discussed.

**Theorem 7.** ∃ an infinite number of irreversible heat pathways $P'_{AB}$ arbitrarily close to a reversible one $P_{AB}$ due to augmenting variables $\Delta$.

**Proof.** Consider two diathermal pathways forming a closed loop in the sequence $A \to B \to A$ along $P_{AB}$ and then back along the irreversible segment $P'_{BA}$. From Axiom 2, the loop is a Riemann sum of $j$ open Carnot cycles, where a typical member is located about points $\{a_1, b_1, b_2, a_2\}$, where segments $(a_1, b_1) \in P_{AB}$, $(b_2, a_2) \in P'_{AB}$ and where segments $(b_1, b_2) \in b'$ and $(a_1, a_2) \in a'$ are adiabatic. Let the increment of heat absorbed along $(a_1, b_1)$ be $Q_{1,[q]}$ at temperature $T_1$ and the heat absorbed (which has an implied negative sign since at this temperature $T_2$ the heat is ejected) about $(b_2, a_2)$ be $Q_{2,[q]}$ for the forms [q] with associated entropy $\mathscr{S}_{[q]}$. When $q = dia$, the system refers to diathermal heat exchange about the paths $P_{AB}$ and $P'_{BA}$ and the work done $W$ given below refers to the work done by the system on the environment, but when $q = tot$ a non-local form of heat transfer is implied, as with the work as given in (1-6) . For what follows, we drop the usage of the [q] subscript notation, for the analysis pertains to both. At optimal efficiency $op$, $Q_{1,op} + Q_{2,op} = -W_{op}$, where $W_{op}$ is the total work done on the system about the $j$ cycle $\{a_1, b_1, b_2, a_2\}$ and $W_{op} = Q_1(f(T_1, T_2))$. Since path $(a_1, b_1)$ is reversible, $|Q_2(P'_{BA})|$ about $(a_2, b_2)$ segment is $|Q_2| = Q_1 + W_{op} + \delta(a_2, b_2)$ where $\delta(a_2, b_2) > 0$ is the dissipation function about segment $(a_2, b_2)$ and if an external field **F** is present, then $\delta(a_2, b_2, \mathbf{F}) > 0$. Excluding external forcing conditions (fields are part of the $\Sigma$ coordinates), $\delta$ is not dependent on the reversible $P_{AB}$ pathway, since $|Q_2| = Q_1(\Sigma) + W_{op}(\Sigma)$ or along the path, the differential heat is $dQ_2 = \mathbf{H}(\Sigma).d\mathbf{\Sigma}$ for a reversible system (e.g. for a perfect gas at an isotherm, $dQ_2 = PdV$) whereas $\delta$ is partially dependent on boundary conditions and field gradients; for if it were completely dependent on $\Sigma$ only, then there would be the expression $dQ_{2,irr} = \mathbf{H}(\Sigma).d\mathbf{\Sigma} - \mathbf{\delta'}(\Sigma).d\mathbf{\Sigma} = \mathbf{F}(\Sigma).d\mathbf{\Sigma}$, implying a fixed dissipation amount with no control which would make $\mathbf{F}(\Sigma)$ indistinguishable from $\mathbf{H}(\Sigma)$ for arbitrary $\Sigma$, including when the cycle is reversible. On the other hand, Axiom 4 states $\exists$ a sequence of points about $(a_2, b_2) \subset P_{AB}(\Sigma_2) \subset \Sigma$ ($\Sigma_2$ represents the sequence of points for the $P_{AB}$ path at the lower portion of the $j$ cycle where heat $Q_2$ is dissipated). Thus it follows that $P'_{BA} = \{P_{BA}, \Delta\}$ where $\Delta$ are augmented variables not in $\Sigma$, so that $P'_{BA}$ has the points in $P_{BA}$ as a subsequence of its entire set.

**Corollary 1**. It is impossible for any irreversible pathway $P'_{BA}$ to contain the same sequence of points as $P_{BA}$ for any path $P_{BA}$.

**Remark 2**. Theorem 7 contradicts the PLE or "local equilibrium hypothesis" as defined here since $P_{AB} \subset \Sigma$ exists for the non-equilibrium sequence, but $\exists$ also the $\Delta$ variables, where $\Delta \notin \Sigma$. Theory does not give the explicit form for $\Delta$, but physical considerations suggests that $\Delta$ includes gradients of the $\Sigma$ coordinates with respect to the space and time coordinates.

2.3 Non-equilibrium molecular dynamics (NEMD) verification of theorems

A simple dimer reaction $2A \leftrightarrow A_2$ which has been previously described for a simulation under low temperature molecular breakdown [8] was simulated this time under a high temperature gradient with and without forced formation of bonds at one end of the MD cell, where the $x$ axis of a rectangular cell was divided into layers where the temperature gradient was applied; the orthogonal $z$ and $y$ axis were 1/16 the length of the unit $x$ axis. In the simulation, free atoms $A$ intereact with other particles (whether $A$ or $A_2$) via a Lennard-Jones spline potential $u_{LJ}$ for distance $r$ where $u_{LJ} = 4\varepsilon\left[\left(\dfrac{\sigma}{r}\right)^{12} - \left(\dfrac{\sigma}{r}\right)^{6}\right]$ for $r \leq r_s$,

$u_{LJ} = a_{ij}(r - r_c)^2 + b_{ij}(r - r_c)^3$ for $r_s \leq r \leq r_c$ and $u_{LJ} = 0$ for $r > r_c$ where $r_s = (26/7)^{\frac{1}{6}}\sigma$ and the cut-off radius is $r_c = (67/48)r_s$. The sum of particle diameters is $\sigma$ and $\varepsilon$ is the potential depth of the spine potential set to unity (LJ units are used throughout where $\sigma = \varepsilon = 1$). The constants $a_{ij}$ and $b_{ij}$ are $a_{ij} = -(24192/3211)\varepsilon/r_s^2$, $b_{ij} = -(387072/61009)\varepsilon/r_s^3$. Any two unbounded atoms interact with the above $u_{LJ}$ potential up to distance $r_f$ with energy $\varepsilon = u_{LJ}(r_f)$ when the potential is switched at the cross-over point to the molecular potential given by $u(r) = u_{vib}(r)s(r) + u_{LJ}[1 - s(r)]$ for the interaction potential between the bonded particles constituting the molecule where $u_{vib}(r)$ is the vibrational potential given by

$$u_{vib}(r) = u_0 + \frac{1}{2}k(r-r_0)^2$$ and the switching function $s(r)$ has the form $s(r) = \dfrac{1}{1+\left(\frac{r}{r_{sw}}\right)^n}$ where

$$\begin{cases} s(r) \to 1 & \text{if } r < r_{sw} \\ s(r) \to 0 & \text{for } r > r_{sw}. \end{cases}$$

The switching function is operative up to $r_b$, after which the molecule ceases to exist and interacts with the potential $u_{LJ}$ like free atoms; similarly, bonded atoms interact with other particles, whether bonded or free with the $u_{LJ}$ potential. The point $r_f$ of formation of a molecule corresponds to the intersection of the harmonic $u_{vib}(r)$ and $u_{LJ}$ curves, and their gradients are almost the same at this point; by the Third dynamical law, momentum is always conserved during the crossover despite possible finite changes in the gradient, provided the point of intersection is defined to belong to just one of either the molecular or atomic potential curves and not both. Total energy is conserved since the curves cross, and uncertainties can only be due to the finite time step per cycle in the Verlet leap frog algorithm. The molecule interacts with its partner after formation at $r_f$ with the $u(r)$ potential until breakup at $r_b$, $r_f < r_b$ (two port loop system). The parameters used here are: $u_0 = -10, r_0 = 1.0, k \sim 2446$ (exact value is determined by the other input parameters), $n = 100, r_f = 0.85, r_b = 1.20,$ and $r_{sw} = 1.11$. The potentials illustrate molecular formation as a hysteresis loop; nearly all transition state kinetic theories under the dubious "time-reversibility-invariance" influence [9], model molecular formation and breakdown along a non-loop pathway where $r_f = r_b$, although most complex reactions such as photochemical lasers have loop mechanisms. The potentials used here bear a striking resemblance to the charge neutralization reaction $K^+ + I^- \to K + I$ [Fig.6.60, pp.375-376; 10] except that the discussion does not explicitly mention the crossing over of the KI and $K^+I^-$ potentials at short distances (high energy), although there is reason to suppose such processes can occur, since the KI potential curve exists at small distances well before the crossover point. We conjecture that the reluctance of experts may be due in part to the influence of "time-reversibility" concepts [9] and this simulation serves to both predict and challenge experimentalists to verify the existence of such reaction mechanisms. The current molecule does not violate any dynamical or thermodynamical principle and hence can be used to verify the above theorems. The current work is a recalculation of previous data taking into consideration the very high energy used here that warrants a much smaller time step ($\delta t = 5.0 \times 10^{-5}$) and a new energy-momentum algorithm scheme at the cross-over point $r_f$ and $r_b$ since if a switch is used, there will be a small energy difference at the time of the switch, and similarly for cross-over without switching, the very steep gradients imply overshoots and energy violation if the Verlet algorithm is used. The solution used here is to impose a novel energy–momentum conservation scheme. If $E_p(r)$ is the interparticle potential (energy) and $E_m(r)$ that for the molecule just after the crossover, the algorithm promotes the particles to a molecule and rescales the particle velocities of only the two atoms forming the bond from $\mathbf{v_i}$ to $\mathbf{v'_i}$ ($i = 1, 2$) where $\mathbf{v'_i} = (1+\alpha)\mathbf{v_i} + \boldsymbol{\beta}$ such that energy and momentum is conserved, yielding $\boldsymbol{\beta} = \dfrac{-\alpha(m_1\mathbf{v_1} + m_2\mathbf{v_2})}{(m_1+m_2)}$

(for momentum conservation) and energy conservation implies that $\alpha$ is determined from the quadratic equation $\alpha^2 qa + 2qa\alpha - \Delta = 0$ with $a = (\mathbf{v_1} - \mathbf{v_2})^2$, $q = \dfrac{m_1 m_2}{2(m_1+m_2)}$ and $\Delta = (E_p - E_m)$ where empirically, success in real solutions for $\alpha$ for each instance of molecular formation is 99.9% and 100% for breakdown-where the $\Delta$ value in this instance is very small ($\sim 1.0 \times 10^{-4}$). This new algorithm coupled with shorter time step ensured excellent thermostatting of the system for equilibrium runs without having to thermostat each layer (64 in all) but only the ends of the cell to ensure uniform temperature throughout the cell; there was no reason to use this algorithm in the previous study [8] because a low temperature decay of the molecules only was observed with no molecular formation. To verify Remark 2, it is sufficient to compare any property $P(S_P)$ at equilibrium (dependent only on variables $\Sigma$) to that for a nonequilibrium simulation along a sequence of non-steady state paths, such as that which exists along the $x$ direction of the MD cell (divided into 64 intervals here) for the same minimum set of numerical values $S_P = \{\Sigma_1, \Sigma_2, ....\Sigma_{min}\}$ of the variables in $\Sigma$ which determines a particular property at equilibrium; if there are differences in the values of a particular property P for the same set $S_P$ in the equilibrium and non-equilibrium runs, then the local equilibrium principle is contradicted, in accordance with Theorem 7. We illustrate this for the equilibrium concentration ratio (which reduces to the equilibrium constant at unit activity for A and $A_2$), the net reaction rate $\sigma_r$, and the flux currents $J_A, J_{A_2}$.

2.3.1 Simulation results

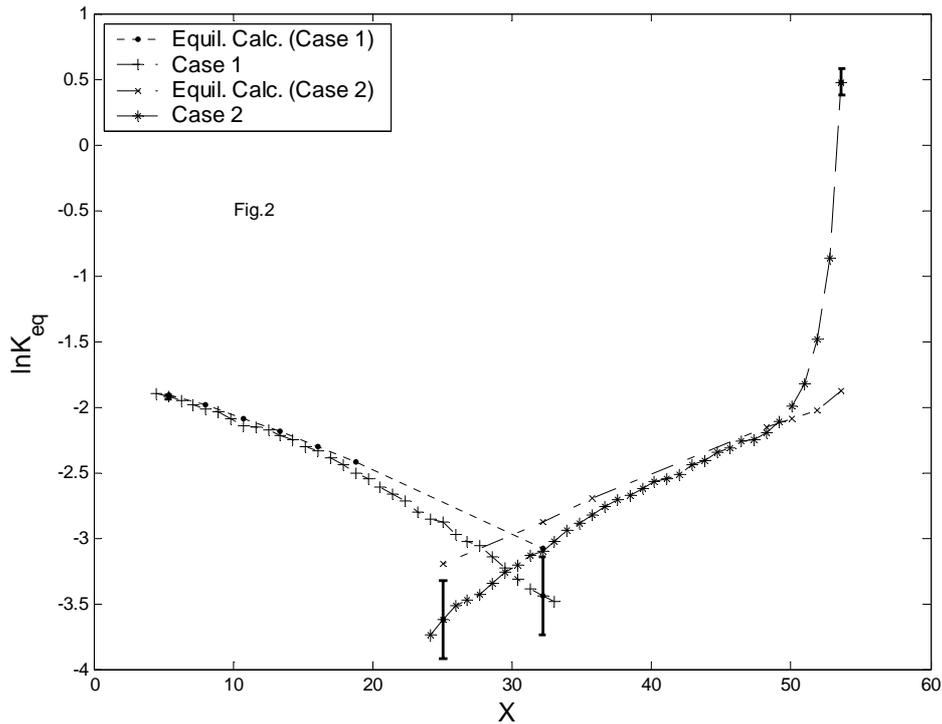

Figure 2. Case 1 and 2 nonequilibrium runs compared with a strictly equilibrium run along X, the distance coordinate from the origin measured from one end of the cell.

Two cases are presented here, Case 1 involves imposing a negative gradient from T=9 to 0.5 across the MD cell, where layers 1-4 ($X < 15$) and 60-64 ($X > 55$) were thermostatted whereas Case 2 had a positive gradient from 0.5 to 9.0, and also had bonds formed at the hot end with rate $81.24 \pm 0.20$. For Case 1 the rate of heat supply at the hot and cold ends (no compensation of the centre-of mass velocity of the MD cell was carried out to prevent drift, which would certainly affect the heat transfer terms) were $(4.12\pm.12)\text{x}10^{-2}$ and $(4.83\pm.58)\text{x}10^{-2}$ respectively, which is equal within the uncertainties (1 standard deviation ). Case 1 was 12M (million) timesteps, 8M at $\delta t = 5.0\times10^{-5}$ and sampling over 4M timesteps, 200 dumps where $\delta t = 5.0\times10^{-4}$. Case 2 was over 18M timesteps, $\delta t = 5.0\times10^{-5}$ for first 14M, and $\delta t = 5.0\times10^{-4}$ for last sampling interval over 4M timesteps. The equilibrium runs (for the same particle densities and temperature as that found in Case 1 and 2 were 6M timesteps, all layers thermostated, and $\delta t = 5.0\times10^{-4}$. Since $-\Delta G^{\varnothing}(T)/(RT) = \ln(K_{eq}\phi(T,\rho))$ with $\phi$ the activity coefficient ratio with limits $\phi \to 1, \rho \to 0$, then $K_{eq} = K_{eq}(T,\rho)$ where $K_{eq} = [A_2][A]^{-2}$, $[X]$ being the number concentration of species $X$ so for the same $(T,\rho)$ variable, $K_{eq}$ is uniquely determined. The thermodynamical equilibrium constant $K_c = K_{eq}\phi(T,\rho)$ would also shift if $K_{eq}$ does, at the same $(T,\rho)$ values. Fig.2 shows a clear violation of PLE with respect to $K_{eq}$ ($K_c$) for $X > 50$ (Case 2). Since for thermodynamical equilibrium $K_c = K_c(\Sigma)$, it follows that the $\Delta$ variables must be evoked to account for the shift.

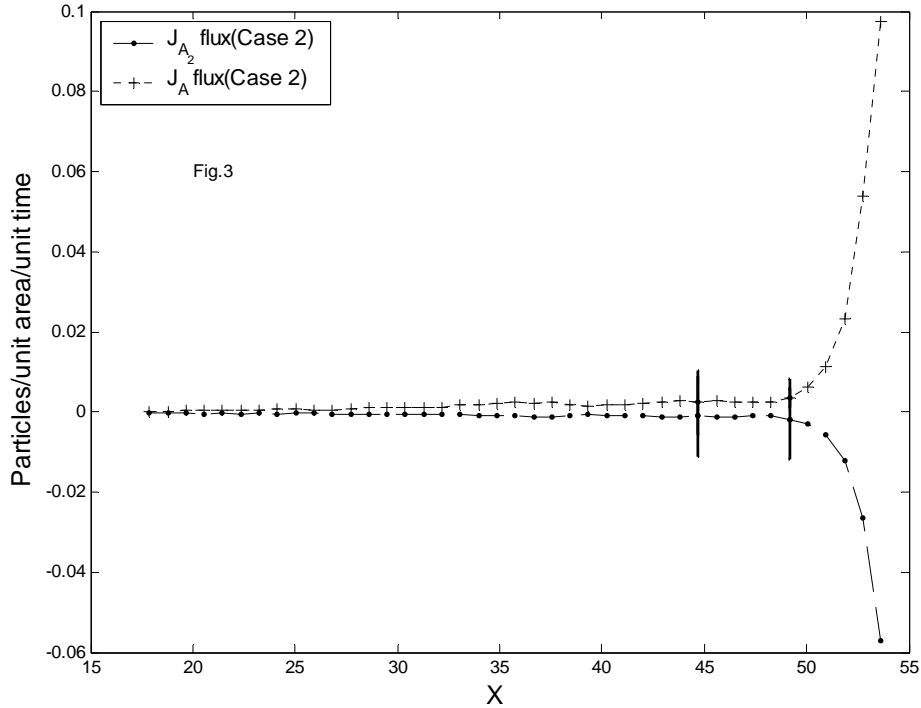

Figure 3. Presence of a bi-directional compensating flux observed at one edge of the cell.

Fig. 3 shows fluxes (very faintly observed in Case 1, but the effect is less than the uncertainty of error, and although the result is not random and is also systematic, it will not be presented since it may still be an artifact of the errors due to simulation) that are bidirectional and compensating starting at approximately $X > 30$, but which becomes acute near the molecular sources located at the high temperature reservoir ($50 < X < 55$). Since dimers are formed (by a hypothetical catalyst) if they are within the range of molecular internuclear distances, there is a constant replenishment of the monomer, leading to a positive flux; the excess production leads to a negative flux into the reactor length, both by diffusion. At low temperatures ($<4$) dimers cannot form, leading to the attenuation at low T. With stationary sources and sinks $\sigma$ ($\sigma_f$ and $\sigma_b$ is the rate of formation and breakdown of dimer in unit time and volume respectively), the conservation equations reads respectively $dc_{A_2}/dt = -\nabla . J_{A_2} + \sigma_f - \sigma_b$ and $dc_A/dt = -\nabla . J_A - 2\sigma_f + 2\sigma_b$ where the $c$'s are the concentrations. The steady state condition is $\nabla . J_A = -2(\sigma_f - \sigma_b) = -2\sigma_r$ and $\nabla . J_{A_2} = \sigma_r$ with $(\sigma_f - \sigma_b) = \sigma_r$; $\sigma_r$ is a scalar flux [p.57-58, 6 a; p.36, 6 b] and at thermodynamical equilibrium $\sigma_r = 0$. If PLE were valid, the $J_A, J_{A_2}$ fluxes would vanish; clearly here, this is not the case, and the second order effect must be related to the $\Delta$ variables; the flux current persists even in the region where PLE holds for $K_e$ (Fig.2). Fig. 4 (line (b)) shows clear violation of this result. To check for flux conservation - so that artifacts can be ruled out - the divergence term is discretized by integration over one layer, using the trapezoidal rule, where $\int_{i-1}^{i} \nabla . J_{A_2} dV = \frac{(\sigma_r(i) - \sigma_r(i-1))\Delta V}{2} = J_{A_2,dif}(i) = J_{A_2}(i) - J_{A_2}(i-1)$ the layer having volume $\Delta V$.

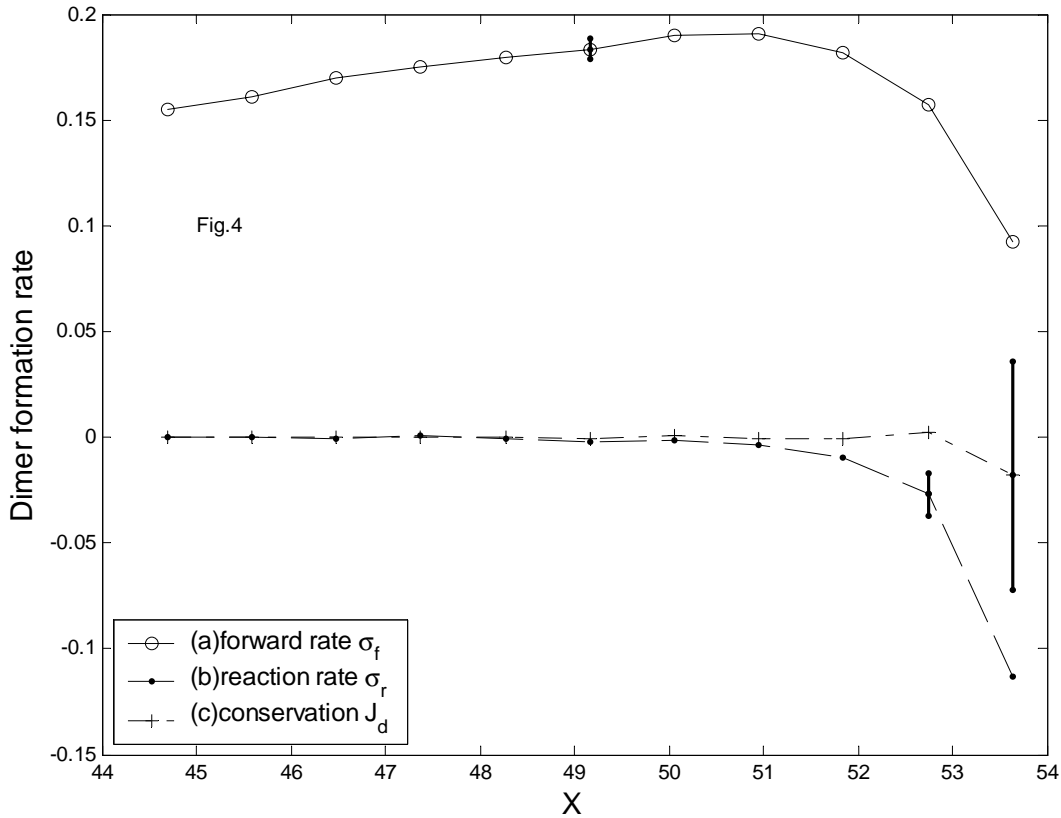

Figure 4. Conservation of matter (via vector divergence) obeyed but reaction rate $\sigma_r \neq 0$.

Similarly, $J_{A,dif}(i) = J_A(i) - J_A(i-1) = -(\sigma_r(i) + \sigma_r(i-1))\Delta V$ leads to $J_d(i) = 2J_{A_2,dif}(i) + J_{A,dif}(i) = 0$. Fig.4 shows that $J_d$ complies with the conservation law rather well, within statistical error, which becomes large due to the different terms in the summation.

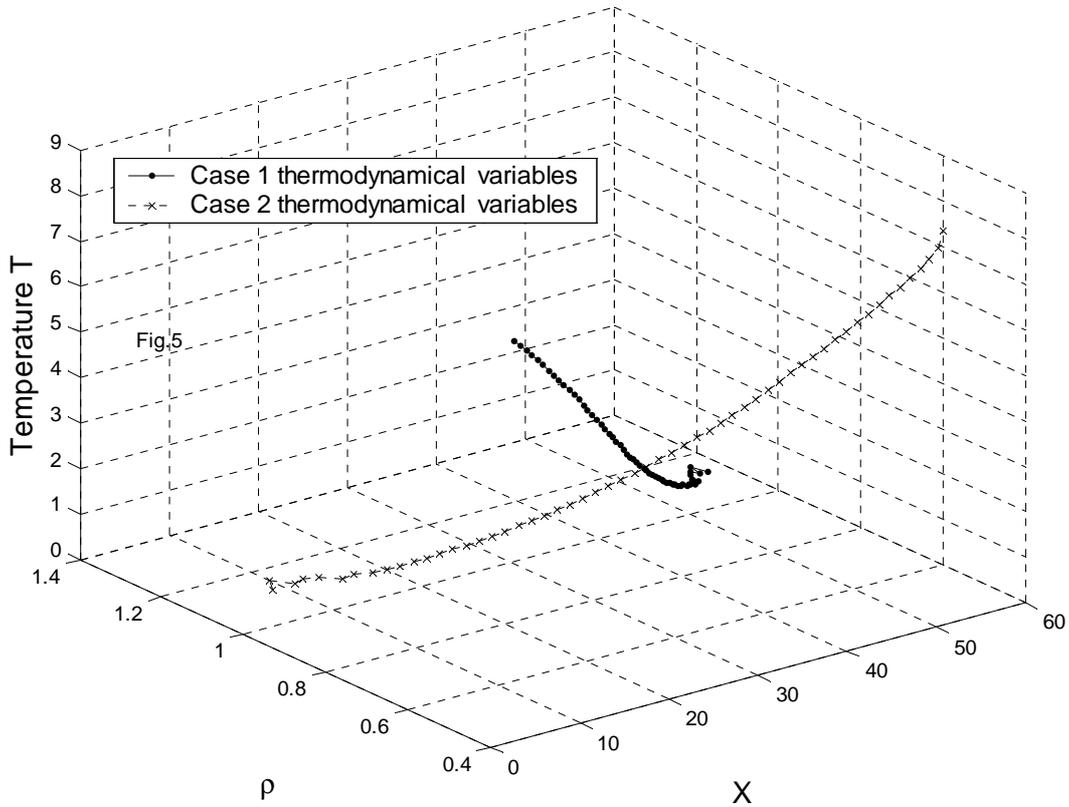

Figure 5. The thermodynamical coordinates for Case1 and 2.

Fig.5 are the thermodynamical coordinates for the entire system, which depicts constant temperatures at the end layers that are thermostated, leading to cluster formation. The simulations show that apart from violations of PLE (supporting Theorem 7) there also exists forces $F_i(\Delta)$ which are vectorial or tensorial functions of $\Delta$ which affects quantities such as $\sigma_r$, a "scalar flux" and $K_{eq}(K_c)$, so that in general $\sigma_r = Q(F_i, \Sigma), K_c = R(F_i, \Sigma)$, so that the Curie principle as commonly understood [pp.36,194,218; 6 b] which states that "in (an isotropic system) there may not be coupling between scalar and vectorial quantities" leading to the setting to zero any cross-coupling coefficients between these terms [eq. 3.19,p.26;6 b] requires scrutiny. (Needless to mention perhaps, it is hard to imagine an isotropic system with long-range non-zero vectorial quantities!). In some treatises [eq.47-48, p.21; 6 c] it is generally stated that "the coupling between any vectorial and scalar process vanishes" due to the principle, which is the basis in constructing equations that couple/decouple flow quantities for non-equilibrium systems, studies and simulations. In fact the methodology is detailed, refined, entertaining no exceptions [p.5 eqs.14-21,p.33-55; 6 a] when setting to zero these cross-coefficients between vectors and scalars.

In a recent mesoscopic rendering [7 b], the ambiguous Gibbsian equations [1] was utilized to model "far from equilibrium" situations, thereby presuming the validity of PLE in these regimes, in contrast to Theorem 7 and the simulation results. This tendency is affirmed in a related work by the remark: "The surprising finding is that we shall need the assumption of local electrochemical equilibrium in the reaction coordinate space" [p.13471; 7 d]. Indeed, the application is "set up by non-equilibirium thermodynamics" [p.9170; 7 e] where "nonequilibrium thermodynamics" specifically refers there to the PLE adhering, linear theory described in [6 a].

## 3. Summary

The above theorems on Clausius inequality and principle of correlation corroborated by NEMD shows that $\Delta$ variables must feature in any reasonable and logical far-from-equilibrium theory. The century of research which decoupled scalar fluxes and vectorial or tensorial forces due to the Curie principle must be revised for far-from-equilibrium theories where the PLE is no longer valid. A novel hysteresis-loop reaction model awaiting experimental verification was employed for the NEMD portion, which employed a new switching algorithm with energy-momentum conservation which can be used to model more conventional molecules.

**Acnowledgements:** The author is grateful to Somsak Sriprayoonsakul, Sugree Phatanapherom and Putchong Uthayopas (H.P.C.N.C, Kasetsart Univ., Thailand) for assistance and support of the cluster computer used here, and Malaysian I.R.P.A. grant (09-02-03-EA0151) for financial assistance for this project.

**References**
[1] C. G. Jesudason, Analysis of open system Carnot cycle and state functions, Nonlinear Anal. Real World Appl., **5** (4), 695-710, 2004.
[2] a A.A Bhalekar, J. Non-Equilb. Thermodyn., **21** 330-338 (1996); b W Muschik, S Gumbel J. Non-Equilb. Thermodyn., **24**(1), 97-106 (1999)
[3] a B.C. Eu, Entropy for irreversible processes, Chem. Phys. Lett., **143**(1), 65-70, (1988); b J. Phys. Chem. **91**(5) 1184-1199 (1987); c Phys. Rev. E (Brief Report) **51**(1) 768-771 (1995); d J. Chem. Phys. **102** 7169-7179 (1995); e J.Chem. Phys. **103**(24) 10652-10662 (1995); f J.Chem. Phys. **104**(3) 1105-1110 (1996); g Phys. Rev. E **54**(3) 2501-2512 (1996); h J. Chem. Phys. **107**(1-4) 222-236 (1997); i J. Phys. Chem. B **103**(40) 8583-8594 (1999); j Kinetic Theory and Irreversible thermodynamics (Wiley-Interscience,New York, 1992)
[4] a M. Chen, B.C. Eu, J Math. Phys **34** 3012-3029 (1993); b M Chen, H.C. Tseng, J. Math. Phys. **39**(1) 329-344 (1998); c D. Jou, J. Casas-Vazquez, G Lebon, Extended Irreversible Thermodynamics, (Springer, Heidelberg,1993); d I Müller,T. Ruggeri, Extended Thermodynamics (Springer, Heidelberg,1993)
[5] C.G. Jesudason, Some consequences of an analysis of the Kelvin-Clausius entropy formulation based on traditional axiomatics, Entropy, **5**, 252-270, (2003)
[6] a S.R. de Groot, P. Mazur, Non-Equilibrium Thermodynamics, (Dover Publ., New York,1982); b K. S. Førland, T. Førland, S.K. Ratkje, Irreversible thermodynamics, (John-Wiley,1988); c W. Yourgrau,A van der Merwe, G. Raw, Treatise on Irreversible and Statistical Thermophysics, (Dover Publ., New York, 1982)
[7] a B. Hafskjold, S. Kjelstrup Ratkje, J. Stat. Phys **78**(1-2) 463-494 (1995); b J.M.G. Vilar, J.M. Rubi, Thermodynamics "beyond" local equilibrium, PNAS **98**(20) 11081-11084 (2001); c D. Bedeaux, S. Kjelstrup, Chem. Eng. Sci., **59**(1) 109-118 (2004); d J.M. Rubi, S. Kjelstrup, J. Phys. Chem. B **107**(48) 13471-13477 (2003); e D. Bedeaux, S.Kjelstrup, J.M. Rubi, J. Chem. Phys. **119**(17),9163-9170 (2003)
[8] C.G. Jesudason, Jurnal Fizik Malaysia, Non-equilibrium molecular dynamics of model chemical reaction in a steady state with breakdown of the local equilibrium hypothesis along the reactor volume , **24**(2) 85-93 (2003)
[9] C. G. Jesudason, I. Time's Arrow, detail balance, Onsager reciprocity and mechanical reversibility. Basic Considerations. *Apeiron*, (http:www.redshift.vif.com) **6(1)** (1999) p.9-23
[10] R.D. Levine and R.B. Bernstein, Molecular Reaction dynamics, (Oxford University Press, England, 1987)